\documentclass[aps,twocolumn,groupedaddress,superscriptaddress,floatfix,prl]{revtex4-2}
\usepackage{amsmath}
\usepackage{mathtools}
\usepackage{amssymb}
\usepackage{graphicx}
\usepackage{textcomp}
\usepackage{bm}	
\usepackage{soul}

\usepackage[usenames,dvipsnames]{color}

\usepackage[braket, qm]{qcircuit}

\begin{document}

\title{Vacuum enhanced charging of a quantum battery}

\author{Tiago F. F. Santos}
\affiliation{Instituto de F\'isica, Universidade Federal do Rio de Janeiro, CP68528, Rio de Janeiro, Rio de Janeiro 21941-972, Brazil}

\author{Yohan Vianna de Almeida}
\affiliation{Instituto de F\'isica, Universidade Federal do Rio de Janeiro, CP68528, Rio de Janeiro, Rio de Janeiro 21941-972, Brazil}

\author{Marcelo F. Santos}\email{Corresponding author: mfsantos@if.ufrj.br}
\affiliation{Instituto de F\'isica, Universidade Federal do Rio de Janeiro, CP68528, Rio de Janeiro, Rio de Janeiro 21941-972, Brazil}

\date{\today}
\begin{abstract}
Quantum batteries are quantum systems that store energy which can then be used for quantum tasks. One relevant question about such systems concerns the differences and eventual advantages over their classical counterparts, whether in the efficiency of the energy transference,  input power, total stored energy or other relevant physical quantities. Here, we show how a purely quantum effect related to the vacuum of the electromagnetic field can enhance the charging of a quantum battery. In particular, we demonstrate how an anti-Jaynes Cummings interaction derived from an off-resonant Raman configuration can be used to increase the stored energy of an effective two-level atom when compared to its classically driven counterpart, eventually achieving full charging of the battery with zero entropic cost.
\end{abstract}

\maketitle
The quest for advanced quantum technologies or the irreversible role of measurements in quantum dynamics are examples of subjects that have stimulated the study of thermodynamics in the microscopic world. An important recent topic of investigation involves the role played by quantum resources in the storage and use of energy by quantized systems~\cite{karen2013, Binder2015, Ferraro2018, Andolina2019, Farina2018, Farina2019, Barra2019, Santos2019, cres2020, sergi2020, Tacchino2020, M2020, Kamin2020, tro2021, tffs2021, FBarra2022, Buy2022, tffs2022, ahmadi2022}. For example, coherence and entanglement have been proven useful to speed up or to super-extend the charging of quantum batteries~\cite{JG2017, Le2018, yuyu2019, zak2021, ghosh2021, Seah2021, BM2020, liu2021}. Experimental results have also shown advances towards the production of microscopic quantum thermal machines and quantum batteries~\cite{rob2016, Lind2019, Klatzow2019, Horne2020, Quach2022, Chang2022}. Most results regarding quantum properties influencing the performance of quantum  batteries, however, focus on increasing the power of the process rather than enhancing the charging capacity. That is because the latter usually requires entropy producing mechanisms~\cite{Barra2019,Santos2019, cres2020, sergi2020, Tacchino2020, M2020, tffs2022} that have deleterious effects in properties such as coherence and entanglement.

In this work we investigate how the quantized nature of part of an entropy preserving charging circuit can influence the charging of a quantum battery. The circuit comprises a classical power source (p.s.) and an auxiliary frequency changer (f.c.). We compare the variation of the internal energy stored in the battery and the efficiency of the work extraction from the p.s., both for a classical and quantum version of the f.c. component. In both cases, the overall dynamics is unitary and, therefore, comes at zero entropic cost. In the classical scenario, both p.s. and f.c. are connected to the battery for a fixed amount of time, $\tau_c$ (``c'' for  classical), unitarily charging its initially thermal state: $\rho_B(\tau_c) = U_c(\tau_c)\rho_B^TU_c^{-1}(\tau_c)$, where $U_c(\tau_c)$ is derived from the coupling Hamiltonian $H_c=H_{B0}+V_{p.s.}(t)+V_{f.c.}(t)$, $V_j(t)$ is the potential created by the circuit component $j$ and $H_{B0}$ is the free Hamiltonian of the battery. Thermal states are free resources in thermodynamics~\cite{Recurso1, Recurso2, Recurso3} and, therefore, ideal to establish the classical benchmark to be challenged by the quantum version. The charging is measured by the variation $\Delta \mathcal{U}$ of internal energy of the battery, where $\mathcal{U} = \textrm{Tr}[\rho_BH_{B0}]$. In the quantized version, $V_{f.c.}(t)$ is replaced by the interaction Hamiltonian $H_{B-f.c.}(t)$ and the initial state must include the f.c. system which is also in a thermal state: $\rho(0) = \rho^T_B\otimes \rho^T_{f.c.}$. The variation of energy of the battery is now given by $\Delta \mathcal{U} = \textrm{Tr}\{[\rho_B(\tau_q)-\rho^T_B]H_{B0}\}$ ( ``q'' for quantum) where $\rho_B(\tau_q) = \textrm{Tr}_{f.c.} U_q(\tau_q) \rho(0)U^{-1}(\tau_q)$ and $U_q$ is the time evolution operator obtained from $H_q = H_{B0}+H_{f.c.0}+H_{B-f.c}(t)+V_{p.s.}(t)$. Note that, in both cases we assume isolation from the environment and the charging does not produce any entropy. For completeness, we later add dissipative non-unitary terms to the dynamics to verify how our results are affected by the heat exchanged with surrounding reservoirs.

We investigate the classical protocol in a particular setup where the battery is an oscillating two-level system of frequency $\omega_{eg}$, the p.s. generates an oscillating potential of frequency $\omega_L >\omega_{eg}$ and the f.c.. generates another potential of frequency $\omega_q = \omega_L - \omega_{eg}$. This situation is commonly found in many different quantum optical experiments~\cite{EIT, CPTr, CCAP, STIRAP, TAI, Wineland2003, Polzik}, where the battery consists of two non-degenerate ground states $\{|g\rangle,|e\rangle\}$ ($\omega_{eg}\equiv \omega_e-\omega_g>0$) of a real or artificial atom and two modes of the electromagnetic field play the role of power supply and f.c.. The couplings are intermediated by a third atomic level $|m\rangle$ working as an ancilla as depicted in Fig. (1a).  Level $|m\rangle$ should only contribute virtually to the transference of energy and has to be adiabatically eliminated from the dynamics. This is achieved when each of p.s. and f.c. couples off-resonantly one of the lower levels of the battery to $|m\rangle$ in a Raman configuration, where $H_{B0} = \hbar \sum_{j=g,e,m} \omega_j \sigma_{jj}$ ($\sigma_{jk} \equiv |j\rangle \langle k|$), $V_{f.c.}(t)=\hbar \Omega_q (\sigma_{em}e^{i\omega_q t}+\sigma_{me}e^{-i\omega_qt})$ and $V_{p.s.}(t) = \hbar \Omega_L (\sigma_{gm}e^{i\omega_L t}+\sigma_{mg}e^{-i\omega_Lt})$. If $\Delta = \omega_{mg}-\omega_L = \omega_{me}-\omega_q \gg \Omega_L,\Omega_q$, the corresponding time evolution $U_c(t)$ induces Rabi oscillations between levels $|g\rangle$ and $|e\rangle$ that are equivalent to directly coupling them through one effective classical field of coupling strength $\bar{\Omega} = \frac{\Omega_L\Omega_q}{\Delta}$~\cite{Molmer2007}. The optimal charging of the battery is then obtained for a full Rabi flip that swaps the populations $p^T_{g,e}$ in the original thermal state $\rho^T_B=\sum_{j=g,e} p^T_j \sigma_{jj}$, where $p^T_{j}=\frac{e^{\frac{\hbar \omega_{j}}{K_BT}}}{Z_B}$ and $Z_B=\sum_{j}e^{\frac{\hbar \omega_{j}}{K_BT}}$. In this case, $\Delta \mathcal{U}^c= \hbar \omega_{eg} [p_g^T-p_e^T]$. Note that this is the most that a unitary transformation can charge an initially thermalized two-level battery and corresponds to the ergotropy $\mathcal {E}^c$ of the resulting state, $\rho_{B}(\tau_c) = p^T_e\sigma_{gg}+p^T_g\sigma_{ee}$. Ergotropy is defined as $\mathcal{E}_{\rho(\tau)}= \sum_{k,j} r_k E_j (|\langle r_k|E_j \rangle|^2 - \delta_{kj})$, where $E_j$ are the eigenenergies of $H_0$ in increasing magnitude, i.e., $E_i \geq E_j$ for $i>j$, and $r_k$ are the eigenvalues of $\rho(\tau)$ in decreasing order, i.e., $r_i \leq r_j$ for $i>j$~\cite{Erg}.

\begin{figure}[h!]
\begin{center}
\includegraphics[width=8cm]{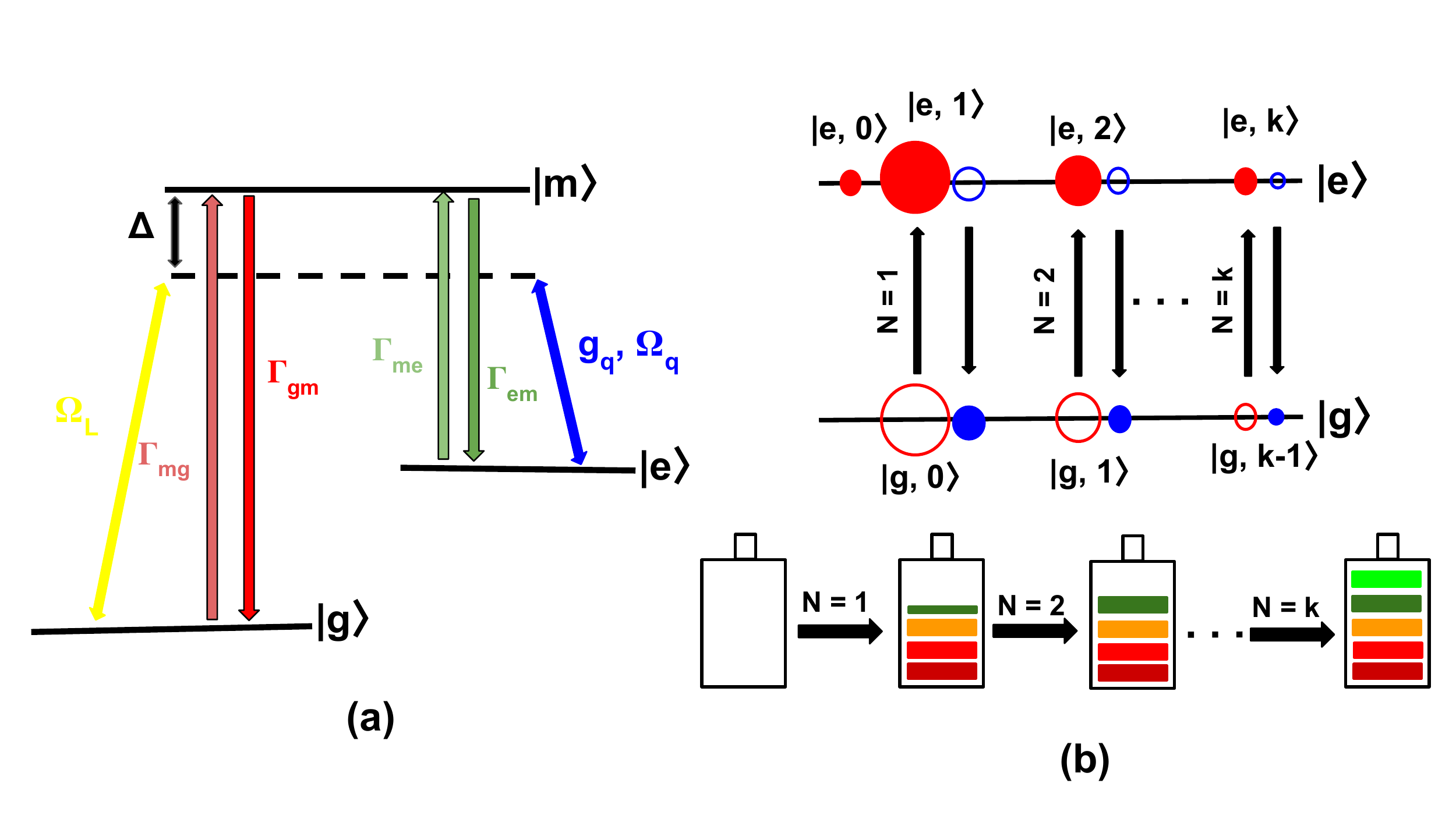}
\caption{(a) off-resonant Raman configuration: the battery is a two-level atom ($\{|g\rangle,|e\rangle\}$); the p.s. is a laser of frequency $\omega_L$ (coupling $\Omega_L$), the f.c. is another harmonic oscillator of frequency $\omega_q$ and couplings $\Omega_q$ (classical) and $g_q$ (quantum). Level $|m\rangle$ is an ancilla that intermediates both couplings. Each channel can also exchange heat with the surrounding reservoirs.the battery. (b) Selective scheme to charge the battery: in each step $N$, a selective Rabi flip transfers energy from $|g, N-1\rangle$ to $|e, N\rangle$.}
\end{center}
\end{figure}

If, now, the classical f.c. is replaced by a quantized field, we need to add its free energy $H_{f.c.0}=\hbar \omega_q \hat{b}^\dagger \hat{b}$ to the Hamiltonian, where $\hat{b}^\dagger$ creates an excitation, and replace $V_{f.c.}(t)$ by the interaction term $H_{B-f.c.
}=\hbar g_q (\sigma_{em}\hat{b}^\dagger+\sigma_{me}\hat{b})$. Once again, for $\Delta \gg \Omega_L,g_q$, we eliminate level $|m\rangle$ and, as shown in~\cite{MKR, MFS,Nicim}, the dynamics of the Battery-f.c. system becomes approximately given by the effective Hamiltonian ($\hbar = 1$)
\begin{eqnarray}
H_{eff} = - \frac{g_q^2 N}{\Delta}\sigma_{gg}-\frac{g_q^2\hat{b}^\dagger \hat{b}}{\Delta}\sigma_{ee}
+ \frac{\Omega_L g_q}{\Delta}(\sigma_{ge}\hat{b}+\sigma_{eg}\hat{b}^\dagger), \nonumber \\
\label{Heff}
\end{eqnarray}
Note that $H_{eff}$ also includes a small correction to the energy difference between levels $|g\rangle$ and $|e\rangle$, given by $\hbar \Delta^{N}_{eg} =\hbar \frac{\Omega^2_L-g^2_q N}{\Delta}$. This term, of the same order of $H_{eff}$, does not affect the conditions for eliminating $|m\rangle$ and can be physically implemented by applying a d.c. Stark shift to the atom.

There are a few aspects of $H_{eff}$ useful for us: first, the a.c. Stark shift correction to level $|e\rangle$ depends on the number of excitations of the f.c. and $|e,0\rangle$ is an eigenstate of $H_{eff}$ with eigenvalue $0$; second, the Rabi oscillations occur in the joint Hilbert space of atom and f.c., splitting it into doublets $\{|g,n\rangle, |e,n+1\rangle\}$. This corresponds to the anti-Jaynes-Cummings (anti-JC) configuration where the p.s. excites both the battery and the f.c. at the same time. Third, each doublet oscillates at its own Rabi frequency given by $\Omega_n=\sqrt{\Delta^{2}_{n}/4+G^2_n}$, where $\Delta_{n}=\frac{r^2 \Omega_L^2 (n+1-N)}{\Delta}$, $G_n = \frac{r\Omega^2_L\sqrt{n+1}}{\Delta}$ and $r\equiv \frac{g_q}{\Omega_L}$, i.e. each doublet is detuned from resonance by an amount $\Delta_n$ proportional to the number of excitations of the f.c..

Such Hamiltonians were predicted and implemented in trapped ions, cavity QED and superconducting circuits, and for $r \gg 1$, they operate in a selective regime where $\Delta_n \gg G_n$ and the Rabi oscillation in all the doublets is highly detuned except if $n=N-1$. In this case, $\{|g,N-1\rangle, |e,N\rangle\}$ oscillates resonantly ($\Delta_{N-1}=0$, $\Omega_{N-1}=\frac{r\Omega_L^2\sqrt{N}}{\Delta}$). Therefore, by properly choosing $\Delta^N_{eg}$ the battery population exchange is conditioned on the number of excitations of the f.c. field as shown in~\cite{MKR, MFS}. For example, for $N =1$, after an interaction time $\tau_q = \frac{\pi \Delta}{2r\Omega_L^2}$, the population in the $\{|g,0\rangle,|e,1\rangle\}$ subspace swaps while all other states only gain number dependent phases. That takes the initial state $\rho(0) =\rho^T_{B}\otimes \rho^T_{f.c.}$ to
\begin{eqnarray}
\rho(\tau_q) &=& p^T_ep^T_0|e,0\rangle\langle e,0|+ p^T_gp^T_0|e,1\rangle\langle e,1| + p^T_ep^T_1|g,0\rangle\langle g,0| \nonumber \\ &+& p^T_gp^T_1|g,1\rangle \langle g,1|+(\sum_{n>1}p^T_n|n\rangle\langle n|)\otimes\rho^T_{B}.
\label{state1}
\end{eqnarray}
Here, $\rho^T_{f.c.}=\sum_n p_n^T \sigma_{nn}$, $p^T_n= e^{-\frac{n\hbar \omega_q}{K_BT}}(1- e^{-\frac{\hbar \omega_q}{K_BT}})$. A simple algebraic manipulation shows that this swap increases the charge of the battery by $\Delta \mathcal{U}^q = (p^T_0p^T_g-p^T_ep^T_1)\hbar \omega_{eg}$. In this case, there is an advantage over $\Delta \mathcal{U}^c$ if $ \frac{p^T_e}{p^T_g}>\frac{1-p^T_0}{1-p^T_1}$. We can better understand this condition at low temperatures. When $K_B T \ll \hbar \omega_q,\hbar \omega_m$, the probabilities $p^T_n$ are negligible for $n>1$ and so is $p^T_m$ and we can approximate $1-p^T_1\approx p^T_0$ and $p^T_e \approx 1-p^T_g$, meaning that $\Delta \mathcal{U}^q>\Delta \mathcal{U}^c$ if $\frac{p^T_e p^T_0}{p^T_g p^T_1} \approx e^{\frac{\hbar \omega_{eg} (\xi-1)}{K_BT}} >1$, where $\xi = \frac{\omega_q}{\omega_{eg}}$. This happens whenever $\xi >1$, i.e. whenever the battery's gap is smaller than one excitation of field $\hat{b}$. In principle, the larger the value of $\xi$, the more accentuated the enhancement due to the vacuum of field $\hat{b}$. This is a purely quantum effect due solely to the vacuum of the f.c. component.

Note, however, that the quantum protocol allows for the relaxation of the $\xi>1$ condition and an even more enhanced charging, which is a much more powerful result, due to the selectivity of $H_{eff}$. In fact, similar Rabi flips can be sequentially applied, each one tuned to resonance by adjusting $\Delta_{eg}^N$ in consecutive subspaces ($N=2,3,...$) as pictorially shown in Fig. (1b). In principle, this sequence must be infinite to maximize the charging of the battery but, in practice, $p^T_n$ tends rapidly to zero unless $T$ is very high, and only a few cycles are required to approach maximum charging. After the sequence, the final state reads $\rho(\sum_j \tau_{qj}) \approx [p^T_e(1-p^T_0)\sigma_{gg}+ (p^T_g+p^T_ep^T_0)\sigma_{ee}$ and the variation of internal energy is $\Delta \mathcal{U}^q = \Delta \mathcal{U}^c+p^T_ep^T_0\hbar \omega_{eg} \geq \Delta \mathcal{U}^c $. This shows an advantage for any positive temperature and independent of $\xi$. More than that, in the limit of $\hbar \omega_q \gg K_B T$, $p^T_0 \rightarrow 1$ and the quantized protocol fully charges the battery, independent of its initial state. This is a purely quantum effect due to the vacuum of the f.c. and consists in the main result of this paper. Not that similar charging can be obtained with open system entropy producing dynamics, such as optical pumping. Here, we match it in an entropy preserving protocol.

This sequence of cycles, however, can be cumbersome and, in practice, escape from the isentropic condition of no heat exchanged with external reservoirs. Furthermore, the classical protocol is much faster, only requiring one Rabi flip. One may wonder, then, if the quantized advantage still holds under equivalent restrictions. To analyze this, we compute, from now on, single shot scenarios designed with a sole detuning adjustment. The energy variation is obtained by solving the Von-Neumann equation with $H_{eff}$. The separation of $H_{eff}$ in doublets makes it easy to derive the time evolution of the eigenstates of $H_{B0}+H_{f.c.0}$. The anti-JC dynamics is similar to the JC and it is simple to show that an initial state $|\Psi(0)\rangle = |g,n\rangle$ evolves to $|\Psi(t)\rangle = e^{-i\Delta_n t/2}[(\cos \Omega_nt+\frac{i\Delta_n}{2\Omega_n} \sin \Omega_nt)|g,n\rangle-\frac{iG_n}{\Omega_n}\sin\Omega_nt|e,n+1\rangle]$. A similar expression can be found for the initial state $|e,n+1\rangle$. Therefore, after evolving for $\tau_q$, the state of the battery changes to $\rho_B(\tau_q) = \textrm{Tr}_{f.c.} [e^{-iH_{eff}\tau_q/\hbar}(\rho^T_{B}\otimes \rho^T_{f.c.})e^{iH_{eff}\tau_q/\hbar}]=\sum p_j \sigma_{jj}$ where $p_g = p_g^T - S(\tau_q)$, $p_e = p_e^T +S(\tau_q)$ and $p_m = p_m^T$ (due to the elimination of level $|m\rangle$). Here, $S(\tau_q) = \sum_{n=0}^{\infty}A_n[p_{g}^Tp_{n}^T(0)-p_{e}^Tp_{n+1}^T]\sin^{2}\left(\frac{\Omega_{n} \tau_q}{2}\right)$, $A_n = \frac{1}{1+\frac{r^2(n+1-N_0)^2}{4(n+1)}}$ and $r=\frac{g_q}{\Omega_L}$ (see Sup. Mat. for full derivation). In this case, $\Delta \mathcal{U}^q= \hbar \omega_{eg}S(\tau_q)$ and the battery's ergotropy reads $\mathcal{E}^q = \hbar \omega_{eg}[p_e^T-p_g^T+2S(\tau_q)] = 2\Delta \mathcal{U}^q - \mathcal{E}^c$. The quantized version will be advantageous whenever $\Delta \mathcal{U}^q >  \mathcal{E}^c$.

A quick inspection of $S(\tau_q)$ shows that, for single shots (ss), it is the non-selective regime of $r \ll 1$ that optimizes the charging of the atom. In this case, all the doublets evolve almost resonantly, each of them contributing to enhance the charge. Because they oscillate at different Rabi frequencies, it is impossible to choose a $\tau_{q,ss}$ that simultaneously maximizes the energy transfer in all of them. The optimal interaction time, which depends on $T$, has to be numerically extracted by maximizing $S(t)$ and, because higher excited states oscillate faster, it gets shorter for higher temperatures. In Fig. (2) we plot the relative gain $K_q \equiv \frac{\Delta \mathcal{U}^q_{ss}-\Delta \mathcal{U}^c}{\Delta \mathcal{U}^c}=\frac{\Delta \mathcal{U}^q_{ss}}{\Delta \mathcal{U}^c}-1$ induced by the single shot quantized protocol as a function of $\xi$ and for two temperatures. Note that, similar to the single shot selective case, $K_q$ increases with $T$ and requires $\xi>1$ to represent positive gain over the classical counterpart.

We also plot in the same figure the efficiency of the work extraction, defined as $\eta \equiv \frac{\mathcal{E}^q}{W_L}$, where $W_L$ is the work injected by the power supply. The first law of thermodynamics says that $W_L = \Delta \mathcal{U}^q+\Delta \mathcal{U}^{fc}$ where $\Delta \mathcal{U}^{fc} = \hbar \omega_q S(\tau_q)$ is the energy variation of the f.c.. Therefore, the efficiency assumes the very simple formula $\eta = \frac{1}{1+\xi} \frac{1+2K_q}{1+K_q}$. For a fixed value of $\xi$, the best efficiency, $\eta = \frac{2}{1+\xi}$, is achieved when $K_q \gg 1$. On the other hand, because $\xi >1$ is a necessary condition for the advantage of the single shot quantum protocol and because $K_q$ increases for larger values of $\xi$, it is clear that the best gains are achieved at lower efficiencies. This should be expected since $\xi \gg 1$ means that most of the energy injected by the power supply is actually going to the f.c.. Note that for each temperature, there is an ideal value of $\xi$ if one wishes for the best gain at a given efficiency.
\begin{figure}[h!]
\begin{center}
\includegraphics[width=8cm]{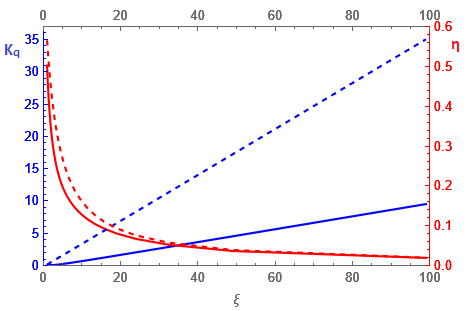}
\caption{Relative gain $K_q = \frac{\Delta \mathcal{U}^q - \Delta \mathcal{U}^c}{\Delta \mathcal{U}^c}$ (blue, straight) and efficiency $\eta=\frac{\mathcal{E}^q}{W_L}$ (red, curved) as a function of parameter $\xi = \frac{\omega_q}{\omega_{eg}}$ for different values of the adimensional temperature $\bar{T} = \frac{K_B T}{\hbar \omega_m}$($\approx$ 0.1 for solid and $\approx$ 0.4 for dashed lines). $\frac{\Delta}{2\pi} = 1$ MHz, $g_q = \frac{\Delta}{600}$, $\Omega_L = \frac{\Delta}{20}$.}
\end{center}
\label{fig2}
\end{figure}

So far, we have considered the isentropic injection of energy by the external source. However, neither the battery nor the f.c. are ever fully isolated from their environment and there will always be heat exchanged with the external reservoir. From the battery's perspective, if both  $|g\rangle \rightarrow |m\rangle$ and $|e\rangle \rightarrow |m\rangle$ transitions are dipole coupled, levels $|g\rangle$ and $|e\rangle$ must be of the same parity and, therefore, cannot be dipole coupled themselves. That means that the time scale for direct energy exchange between them is usually much slower than any other time scale of the problem and, in general, the corresponding heat channel can be ignored. Considering the standard weak coupling to thermal reservoirs, the overall dynamics of the system is, then, governed by a master equation of the form $\dot{\rho} = -\frac{i}{\hbar}[H_q,\rho] +\mathcal{L}(\rho)$~\cite{book}, where $\mathcal{L}(\rho) = \sum_s \Gamma_s [2L_s\rho L_s^\dagger-\{L_s^\dagger L_s,\rho\}]$, with $s=gm,mg,em,me,+,-$. The rates of the non-unitary parts are given by $\Gamma_{jm} = \gamma_{0j}(\bar{n}_j+1)$, $\Gamma_{mj} = \gamma_{0j}\bar{n}_j$, $\Gamma_{-} = \gamma_{0q}(\bar{n}_q+1)$, and $\Gamma_{+} = \gamma_{0q}\bar{n}_q$. Here, the $\gamma_0$'s indicate the spontaneous decay rates and $\bar{n}$'s the average number of photons of the thermal reservoir at frequencies $\omega_{mj}$ and $\omega_q$. The respective jump operators are $L_{jk}=\sigma_{jk}$, $L_{-}=\hat{b}$ and $L_+=\hat{b}^\dagger$.
\begin{figure}[h!]
\begin{center}
\includegraphics[width=8cm]{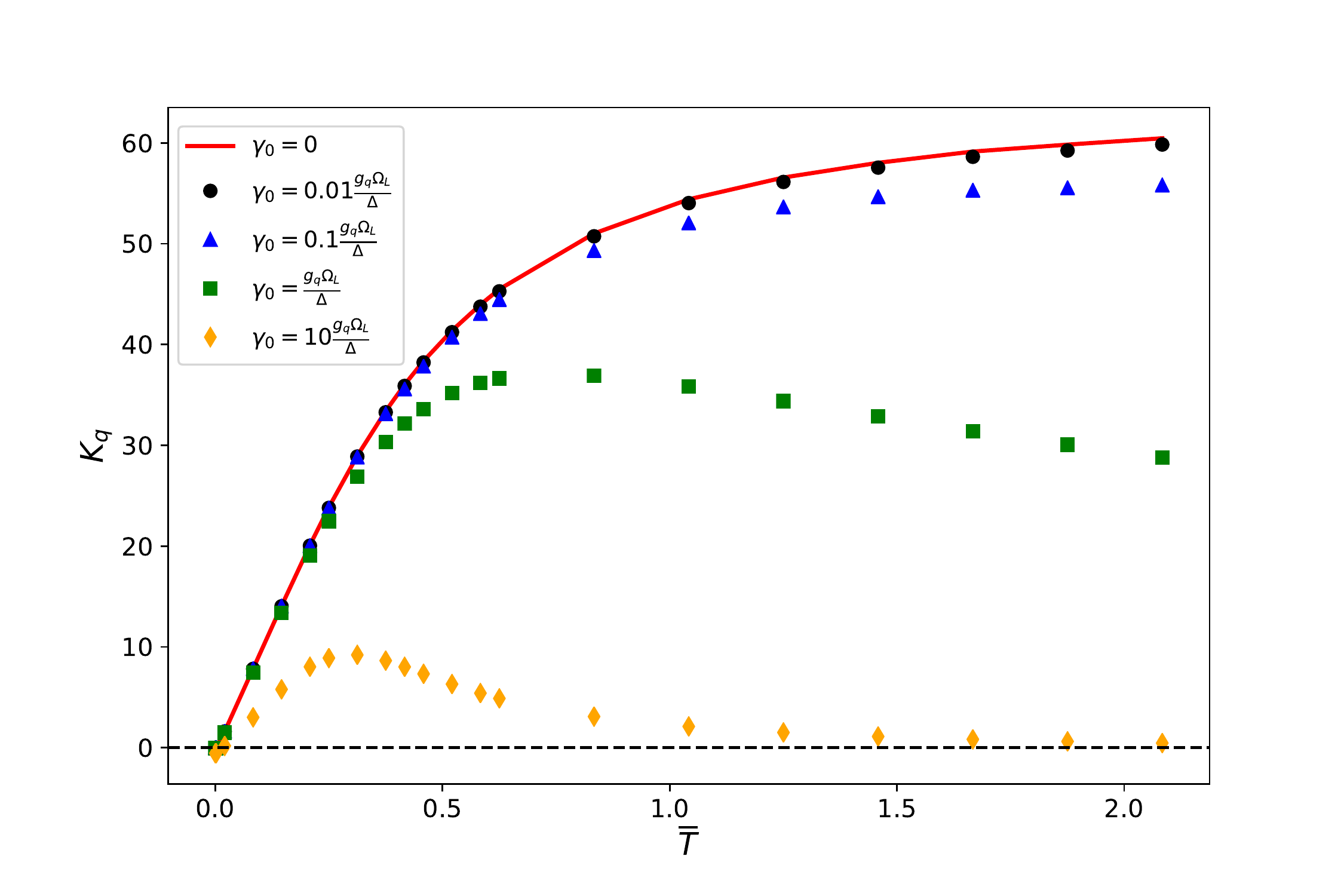}
\caption{Relative gain as a function of the adimensional temperature $\bar{T} \equiv \frac{k_B T}{\hbar \omega_m}$ for different values of spontaneous decay rates $\gamma_0$ and for $\xi = 99$, $\frac{\Delta}{2\pi} = 1$ MHz, $g_q = \frac{\Delta}{600}$, $\Omega_L = \frac{\Delta}{20}$. The solid curve is obtained from the unitary evolution with $H_{eff}$. The dotted curves are numerical solutions of the open system dynamics (master equation) with full Hamiltonian $H_q$.}
\end{center}
\end{figure}

The couplings to the thermal reservoirs establish at least four typical regimes to the problem, depending on their strength. The first one, already addressed, corresponds to $\gamma_0$'s much smaller than the effective coupling $\frac{g_q\Omega_L}{\Delta}$ and $k_BT\ll \hbar \omega_{eg},\hbar \omega_q$. This is well approximated by the isentropic dynamics considered so far. However, we saw that the higher the temperature, the more advantageous the quantum protocol is. This may not hold true when we take into consideration the heat exchanges with the reservoir. As the spontaneous decay rates increase, a combination of effects begin to affect the charging of the battery and may even create optimal temperatures for better quantum gain.

In Fig. (3) we present $K_q$ as a function of the adimensional temperature $\bar{T} \equiv \frac{k_B T}{\hbar \omega_m}$ for different values of $\gamma_0$.  $\bar{T}$ is relevant to the problem because it regulates the population of level $|m\rangle$. Although each reservoir has its own spontaneous decay rate, they all produce similar effects on both $K_q$ and $\eta$, therefore we have considered a single $\gamma_0$ for all of them. The result was obtained by solving the full dynamics of the open quantum system and choosing the best $\tau_{q,ss}$ for each temperature. In these plots, $\frac{\omega_m}{2\pi}=10^{12}Hz$, $\Delta =2\pi MHz= 600 g=20 \Omega_L$, $\xi = 99$, $r =1/30$. As previously discussed, for $\gamma_0 \ll \frac{g_q\Omega_L}{\Delta}$ we reach the unitary regime calculated with Hamiltonian~(\ref{Heff}) (solid curve), except for very high temperatures ($\bar{T} \sim 2$) when the population of level $|m\rangle$ becomes too significant and start to affect the protocol as a whole. As we increase $\gamma_0$, effects such as decoherence of the f.c. field and the augmented relaxation rates $\Gamma_j$ begin to limit the quantum advantage. These effects become particularly relevant when $\Gamma$'s rates approach the effective battery-f.c. coupling $g_q\Omega_L/\Delta$. Note, however, that even for such values of dissipation, the quantum protocol can still produce gains $30$ times larger than its classical counterparts for $\xi = 99$. Finally, a fourth effect takes place for higher values of $\gamma_0$ and at much higher temperatures: when $\Gamma$'s become of the order of $\Delta$ the heat exchange eventually brings the transitions back into resonance in which case level $|m\rangle$ cannot be adiabatically eliminated anymore and the charging scheme breaks down.
\begin{figure}[h!]
\begin{center}
\includegraphics[width=8cm]{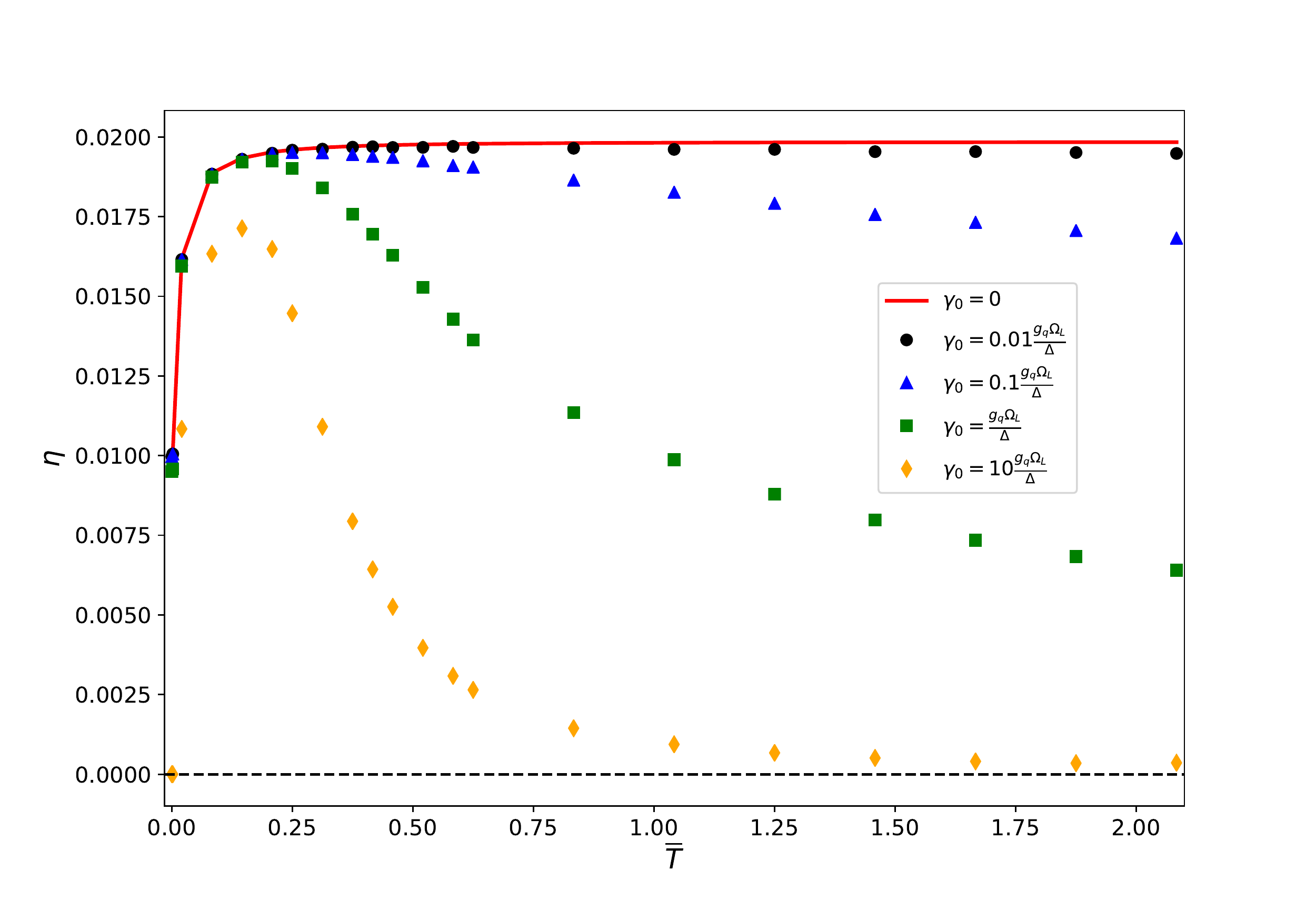}
\caption{Efficiency as a function of the adimensional temperature $\bar{T} \equiv \frac{k_B T}{\hbar \omega_m}$ for different values of spontaneous decay rates $\gamma_0$ and for $\xi = 99$, $\frac{\Delta}{2\pi} = 1$ MHz, $g_q = \frac{\Delta}{600}$, $\Omega_L = \frac{\Delta}{20}$. The solid curve is obtained from the unitary evolution with $H_{eff}$. The dotted curves are numerical solutions of the open system dynamics (master equation) with full Hamiltonian $H_q$.}
\end{center}
\end{figure}

In Fig. (4) we repeat the numerical calculations of the open system dynamics (same parameters), this time for the efficiency. Again, we see that very low $\gamma_0$'s are consistent with the isentropic hypothesis, whereas higher values of the spontaneous decay rates severely affect the efficiency, specially for higher values of $\bar{T}$. Note that for some parameters, the plotted efficiency is corrected to $\eta = \frac{\mathcal{E}^q}{W_L+Q_{em}}$ to adjust for the fact that the $|e\rangle \rightarrow |m\rangle$ reservoir may also inject energy in the system in the form of heat $Q_{em}$. The correction takes place whenever we obtain $Q_{em}>0$.

To conclude, we have shown that the quantized nature of a component of a charging circuit can significantly enhance the isentropic charging of a quantum battery when benchmarked against its classical counterpart. This is a purely quantum effect due to the vacuum state of the quantized component and the ability to selectively manipulate quantum states in the Hilbert space. We have also shown that our protocol can achieve the same full charging capacity of open system entropy producing equivalent schemes. We have demonstrated the effect in a typical setup of off-resonant Raman population transfer in three-level $\lambda-$configuration where the power supply is an external laser field and the quantized component is a harmonic oscillator. This example is particularly useful due to its broad presence in a variety of quantum optical setups such as trapped ions and atoms, cavity QED, superconducting qubits, quantum dots and many other equivalent experiments. 
\acknowledgments
This work was supported by  CNPq Projects 302872/2019-1, INCT-IQ 465469/2014-0, and FAPERJ project E-26/202.576/2019. TFFS and YVA thank Capes for financial support.


\begin{thebibliography}{99}

\bibitem{karen2013} K. V. Hovhannisyan, M. Perarnau-Llobet, M.s Huber, and A. Acín, Entanglement Generation is Not Necessary for Optimal Work Extraction, Phys. Rev. Lett. {\bf 111}, 240401 (2013).

\bibitem{Binder2015} F. C. Binder, S. Vinjanampathy, K. Modi \& J. Goold, Quantacell: power charging of quantum batteries, New Journal of Physics {\bf 17}, 075015 (2015).

\bibitem{Ferraro2018} D. Ferraro, M. Campisi, G. M. Andolina, V. Pellegrini \& M. Polini, High-power collective charging of a solid-state quantum battery, Phys. Rev. Lett. {\bf 120}, 117702 (2018).

\bibitem{Andolina2019} G. M. Andolina, M. Keck, A. Mari, M. Campisi, V. Giovannetti and M. Polini, Extractable Work, the Role of Correlations, and Asymptotic Freedom in Quantum Batteries, Phys. Rev. Lett. {\bf 122}, 047702 (2019).

\bibitem{Farina2018} G. M. Andolina, D. Farina, A. Mari, V. Pellegrini, V. Giovannetti, M. Polini, Charger-mediated energy transfer in exactly solvable models for quantum batteries, Phys.Rev. B {\bf98}, 205423 (2018).

\bibitem{Farina2019} D. Farina, G. M. Andolina, A. Mari, M. Polini and Vittorio Giovannetti, Charger-mediated energy transfer for quantum batteries: An open-system approach, Phys. Rev. B {\bf 99}, 035421 (2019).

\bibitem{Barra2019} F. Barra, Dissipative Charging of a Quantum Battery, Phys. Rev. Lett. {\bf 122}, 210601 (2019).

\bibitem{Santos2019} A.C. Santos, B. Çakmak, S. Campbell and N.T. Zinner, Stable adiabatic quantum batteries, Phys. Rev. E {\bf 100}, 032107 (2019).

\bibitem{cres2020} A. Crescente, M. Carrega, M. Sassetti and D. Ferraro, Charging and energy fluctuations of a driven quantum battery, New J. Phys. {\bf 22} 063057 (2020).

\bibitem{sergi2020} Sergi Julia-F\`arr\'e, Tymoteusz Salamon, Arnau Riera, Manabendra N. Bera, and Maciej Lewenstein, Bounds on the capacity and power of quantum batteries, Phys. Rev. Research {\bf2}, 023113 (2020).

\bibitem{Tacchino2020} F. Tacchino, T. F. F. Santos, D. Gerace, M. Campisi, and M. F. Santos, Charging a quantum battery via nonequilibrium heat current, Phys. Rev. E {\bf 102}, 062133 (2020).

\bibitem{M2020} M. Carrega, A. Crescente, D. Ferraro and M. Sassetti, Dissipative dynamics of an open quantum battery, New J. Phys. {\bf 22} 083085 (2020).

\bibitem{Kamin2020} F. H. Kamin, F. T. Tabesh, S. Salimi, and Alan C. Santos, Entanglement, coherence, and charging process of quantum batteries, Phys. Rev. E {\bf102}, 052109 (2020).

\bibitem{tffs2021} T. F. F. Santos, F. Tacchino, D. Gerace, M. Campisi and M. F. Santos, Maximally efficient quantum thermal machines fueled by nonequilibrium steady states, Phys. Rev. A {\bf 103}, 062225 (2021).

\bibitem{tro2021} Thiago R. de Oliveira and Daniel Jonathan, Efficiency gain and bidirectional operation of quantum engines with decoupled internal levels, Phys. Rev. E {\bf 104}, 044133 (2021).

\bibitem{FBarra2022} Javier Carrasco, Jerónimo R. Maze, Carla Hermann-Avigliano, and Felipe Barra, Collective enhancement in dissipative quantum batteries, Phys. Rev. E {\bf 105}, 064119 (2022).

\bibitem{Buy2022} I. Maillette de Buy Wenniger, S. E. Thomas, M. Maffei, S. C. Wein, M. Pont,  A. Harouri,  A. Lema\^{ı}tre, I. Sagnes, N. Somaschi, A. Auff\`{e}ves,  P. Senellart. Coherence-powered work exchanges between a solid-state qubit and light fields, {\bf arXiv:2202.01109} (2022).

\bibitem{tffs2022} T. F. F. Santos and Marcelo F. Santos, Efficiency of optically pumping a quantum battery and a two-stroke heat engine, {\bf arXiv:2208.02373} (2022), accepted in Phys. Rev. A.

\bibitem{ahmadi2022} B. Ahmadi, P. Mazurek, R. R. Rodríguez, S. Barzanjeh, R. Alicki, and P. Horodecki, Catalysis in Charging Quantum Batteries, {\bf arXiv:2205.05018} (2022).

\bibitem{JG2017} Francesco Campaioli, Felix A. Pollock, Felix C. Binder, Lucas C\'eleri, John Goold, Sai Vinjanampathy, and Kavan Modi, Enhancing the Charging Power of Quantum Batteries, Phys. Rev. Lett. 118, 150601 (2017).

\bibitem{Le2018} T. P. Le, J. Levinsen, K. Modi, M. M. Parish, and F. A. Pollock, Spin-chain model of a many-body quantum battery, Phys. Rev. A {\bf 97}, 022106 (2018).

\bibitem{yuyu2019} YY. Zhang, TR. Lang, L. Fu and X. Wang, Powerful harmonic charging in a quantum battery, Phys. Rev. E {\bf 99}, 052106.

\bibitem{zak2021} S. Zakavati, F. T. Tabesh, and S. Salimi, Bounds on charging power of open quantum batteries, Phys. Rev. E {\bf 104}, 054117 (2021).

\bibitem{ghosh2021} S. Ghosh, T. Chanda, S. Mal, and A. Sen, Fast charging of a quantum battery assisted by noise, Phys. Rev. A {\bf 104}, 032207 (2021).

\bibitem{Seah2021} S.Seah, M. Perarnau-Llobet, G. Haack, N. Brunner, and S. Nimmrichter, Quantum Speed-Up in Collisional Battery Charging, Phys. Rev. Lett. {\bf127}, 100601 (2021).

\bibitem{BM2020} James Q. Quach, and William J. Munro,  Using Dark States to Charge and Stabilize Open Quantum Batteries, Phys. Rev. Applied, 14, 024092 (2020).

\bibitem{liu2021} JX. Liu, HL. Shi, YH. Shi, XH Wang, and WL. Yang, Entanglement and work extraction in the central-spin quantum battery, Phys. Rev. B {\bf 104}, 245418 (2021).

\bibitem{rob2016} J. Ro\ss nagel, S. T. Dawkins, K. N. Tolazzi, O. Abah, E. Lutz, F. Schmidt-Kaler, K. Singer, A single-atom heat engine. Science {\bf 352}, 325–329 (2016).

\bibitem{Lind2019} D. von Lindenfels, O. Gr\"{a}b, C. T. Schmiegelow, V. Kaushal, J. Schulz, M. T. Mitchison, J. Goold, F. Schmidt-Kaler, and U. G. Poschinger, Spin heat engine coupled to a harmonic oscillator flywheel, Phys. Rev. Lett. 123, 080602 (2019).

\bibitem{Klatzow2019} J. Klatzow, J. N. Becker, P. M. Ledingham, C. Weinzetl, K. T. Kaczmarek, D. J. Saunders, J. Nunn, I. A. Walmsley, R. Uzdin, and E. Poem, Experimental demonstration of quantum effects in the operation of microscopic heat engines, Phys. Rev. Lett. {\bf122}, 110601 (2019).

\bibitem{Horne2020} N. Van Horne, D. Yum, T. Dutta, P. H\"{a}nggi, J. Gong, D. Poletti, and M. Mukherjee, Single-atom energy-conversion device with a quantum load, npj Quantum Information {\bf6}, 37 (2020).

\bibitem{Quach2022} James Q. Quach, Kirsty E. McGhee, Lucia Ganzer, Dominic M. Rouse, Brendon W. Lovett, Erik M. Gauger, Jonathan Keeling, Giulio Cerullo, David G. Lidzey, and Tersilla Virgili. Superabsorption in an organic microcavity: Toward a quantum battery. Science advances, {\bf8(2)}:eabk3160, (2022).

\bibitem{Chang2022} Chang-Kang Hu et al, Optimal charging of a superconducting quantum battery, Quantum Sci. Technol. {\bf7} 045018 (2022)

\bibitem{Recurso1} J. Goold, M. Huber, A. Riera, L. del Rio and P. Skrzypczyk, The role of quantum information in thermodynamics—a topical review, J. Phys. A: Math. Theor. {\bf49} 143001 (2016).

\bibitem{Recurso2} Matteo Lostaglio, An introductory review of the resource theory approach to thermodynamics, Rep. Prog. Phys {\bf 82} 114001 (2019).

\bibitem{Recurso3} V. Narasimhachar, S. Assad, F.C. Binder, J. Thompson, B. Yadin, and Mile Gu, Thermodynamic resources in continuous-variable quantum systems, npj Quantum Inf {\bf 7}, 9 (2021).

\bibitem{EIT} Michael Fleischhauer, Atac Imamoglu, and Jonathan P. Marangos, Electromagnetically induced transparency: Optics in coherent media, Rev. Mod. Phys. {\bf77}, 633 (2005)

\bibitem{CPTr} K. Bergmann, H. Theuer, and B. W. Shore, Coherent population transfer among quantum states of atoms and molecules, Rev. Mod. Phys. {\bf70}, 1003 (1998)

\bibitem{CCAP} Petr Kr\'al, Ioannis Thanopulos, and Moshe Shapiro, Colloquium: Coherently controlled adiabatic passage, Rev. Mod. Phys. {\bf79}, 53 (2007)


\bibitem{STIRAP} Nikolay V. Vitanov, Andon A. Rangelov, Bruce W. Shore, and Klaas Bergmann, Stimulated Raman adiabatic passage in physics, chemistry, and beyond, Rev. Mod. Phys. {\bf89}, 015006 (2017)

\bibitem{TAI} Carl E. Wieman, David E. Pritchard, and David J. Wineland, Atom cooling, trapping, and quantum manipulation, Rev. Mod. Phys. {\bf71}, S253 (1999)

\bibitem{Wineland2003} D. Leibfried, R. Blatt, C. Monroe, and D. Wineland, Quantum dynamics of single trapped ions, Rev. Mod. Phys. {\bf75}, 281 (2003)

\bibitem{Polzik} Klemens Hammerer, Anders S. Sørensen, and Eugene S. Polzik, Quantum interface between light and atomic ensembles, Rev. Mod. Phys. {\bf82}, 1041 (2010).

\bibitem{Molmer2007} E Brion, L H Pedersen1 and K Mølmer, J. Phys. A: Math. Theor. {\bf40}, 1033 (2007)

\bibitem{Erg} A. E. Allahverdyan, R. Balian, and Th. M. Nieuwen-huizen, Maximal work extraction from finite quantum systems, Europhys. Lett. {\bf67}, 565 (2004).

\bibitem{MKR} M. F. Santos, E Solano, RL de Matos Filho, Conditional large Fock state preparation and field state reconstruction in Cavity QED, Phys. Rev. Lett. {\bf87}, 93601 (2001).


\bibitem{MFS} MF Santos, Universal and deterministic manipulation of the quantum state of harmonic oscillators: a route to unitary gates for Fock State qubits. Phys. Rev. Lett. {\bf 95}, 010504 (2005).

\bibitem{Nicim} N. Zagury, A. Arag\~ao, J. Casanova, and E. Solano, Unitary expansion of the time evolution operator, Phys. Rev. A, {\bf82}, 042110 (2010)


\bibitem{book} H. P. Breuer and F. Petruccione, \emph{The theory of open quantum systems} (Oxford University Press, 2002).


\end{thebibliography}
\end{document}